\documentclass{aastex}
\usepackage{emulateapj5}

\begin{document}

\title{Are Giant Planets Forming Around HR 4796A?}
\author{C.\ H.\ Chen\altaffilmark{1} \& I. \ Kamp\altaffilmark{2}}
\altaffiltext{1}{National Research Council Resident Research Associate,
       NASA/Jet Propulsion Laboratory, M/S 169-506, 4800 Oak Grove Drive, 
       Pasadena, CA 91109; christine.chen@jpl.nasa.gov}
\altaffiltext{2}{Leiden Observatory, PO Box 9513, 2300 RA Leiden, The 
       Netherlands; kamp@strw.leidenuniv.nl}

\begin{abstract}
We have obtained \emph{FUSE} and \emph{HST} STIS spectra of HR 4796A, a 
nearby 8 Myr old main sequence star that possesses a dusty circumstellar 
disk whose inclination has been constrained from high resolution 
near-infrared observations to be $\sim$17$\arcdeg$ from edge-on. We searched 
for circumstellar absorption in the ground states of \ion{C}{2} 
$\lambda$1036.3, \ion{O}{1} $\lambda$1039.2, \ion{Zn}{2} $\lambda$2026.1, 
Lyman series H$_{2}$, and CO (A-X) and failed to detect any of these species.
We place upper limits on the column densities and infer upper limits on the 
gas masses assuming that the gas is in hydrostatic equilibrium, is 
well-mixed, and has a temperature, $T_{gas}$ $\sim$ 65 K. Our measurements 
suggest that this system possesses very little molecular gas. Therefore, we 
infer an upper limit for the gas:dust ratio ($\leq$4.0) assuming that the gas
is atomic. We measure less gas in this system than is required to form the 
envelope of Jupiter.
\end{abstract}


\section{Introduction}
Planets, asteroids, and comets are believed to form in circumstellar disks
around stars with ages $<$few 100 Myr. Current (uncertain) models suggest
that Jovian planets form either via rapid gravitational collapse through disk
instability within a few hundred years (Boss 2000) or via coagulation of dust 
into solid cores within the first $\sim$1 Myr and accretion of gas into thick
hydrogen atmospheres within the first $\sim$10 Myr (Ruden 1999). At present, 
the  timescales on which giant planets form and accrete their atmospheres 
have not been well constrained observationally. Studies of objects with ages 
$\sim$10 Myr afford the opportunity to constrain the timescale for gas 
dissipation around young stars and therefore the timescale for giant planet 
formation.

Although the circumstellar disks around pre-main sequence stars possess
both gas and dust, the circumstellar disks around main sequence stars
may be gas poor. Millimeter searches for CO around stars with ages 1 - 10 
Myr suggest that within a few Myr, less than a Jupiter mass of molecular gas 
remains in the circumstellar environment (Zuckerman, Forveille, \& Kastner 
1995). However, CO may be depleted in the outer regions of disks because
it freezes out onto dust grains at temperatures, T$_{gas}$ $\sim$ 20 K. Since
gas with interstellar or solar abundance is mainly composed of hydrogen and
H$_{2}$ does not freeze onto grains, recent searches for circumstellar gas 
have focused on H$_{2}$. \emph{ISO} searches for emission from the S(0) and 
S(1) transitions of H$_{2}$ at 28 $\mu$m and 17 $\mu$m respectively, suggest 
that 50 - 2000 $M_{\earth}$ H$_{2}$ exists around $\beta$ Pic, 49 Cet, and HD 
135344 (Thi et al. 2001). However, \emph{FUSE} searches for H$_{2}$ 
absorption in the Lyman band toward $\beta$ Pic place upper limits on the 
column density of H$_{2}$, N(H$_{2}$) $\leq$10$^{18}$ cm$^{-2}$, which is 3 
orders or magnitude lower than that inferred from \emph{ISO} (Lecavelier des 
Etangs et al. 2001). The Thi et al. (2001) results not only apparently 
conflict with the ultraviolet H$_{2}$ measurements but also with millimeter 
emission line studies which suggest that $\beta$ Pic possesses $<$10$^{-5}$ 
M$_{\earth}$ CO (Liseau \& Artymowicz 1998). Ground-based follow-up of 
\emph{ISO} H$_{2}$ detections toward Herbig Ae and T-Tauri stars yielded 
upper limits which were smaller than \emph{ISO} detections (Richter et al. 
2002), suggesting that the H$_{2}$ must lie in an extended cloud. However, an
H$_{2}$ cloud around $\beta$ Pic should have produced a detectable 
ultraviolet absorption line (Jura 2003). Recently, fluorescent H$_{2}$ 
Lyman-band emission has been detected from the surface layers of the 
circumstellar disk around the $\sim$10 Myr old star TW Hydrae (Herczeg et al.
2002) using \emph{HST} STIS. We have carried out a \emph{FUSE} and STIS 
search for circumstellar molecular and atomic gases around the $\sim$8 Myr 
(Stauffer et al. 1995) debris disk object HR 4796A, another member of the 
TW Hydrae Association, to help constrain the gas dissipation timescale.

HR 4796A is a main sequence A0V star with a \emph{Hipparcos} distance of
67.1 pc (see Table 1 for a summary of stellar properties) and a fractional 
dust luminosity L$_{IR}$/L$_{*}$ = 5$\times$10$^{-3}$ (Jura et al. 1998). 
Submillimeter observations suggest that it possesses $\geq$0.25 $M_{\earth}$ 
dust (Greaves et al. 2000), more circumstellar dust than any other main
sequence A-type star in the Yale Bright Star Catalog (Jura et al. 1991). High 
resolution scattered light imaging has resolved a narrow, dusty ring at 70 AU
from HR 4796A with an inclination of $\sim$17$\arcdeg$ from edge-on 
(Schneider et al. 1999). High resolution mid-infrared imaging has resolved 
the disk in thermal emission (Koerner et al. 1998; Jayawardhana et al. 1998).
The presence of a central clearing (Jura et al. 1998) and of a brightness 
asymmetry from the two lobes of the disk (Telesco et al. 2000; Weinberger, 
Becklin, \& Schneider 2000) have led to speculation that the circumstellar 
disk possesses a planetary companion at $\leq$70 AU (Wyatt et al. 1999). 
Blackbody fitting to the \emph{IRAS} fluxes suggests that the grain 
temperature, $T_{dust}$ = 110 K (Jura et al. 1995).

Previous studies of circumstellar gas around HR 4796A have not been very
successful. Searches for millimeter and submillimeter emission from CO 
(J=3$\rightarrow$2, J=2$\rightarrow$1, J=1$\rightarrow$0), used as a tracer 
for H$_{2}$, have failed (Zuckerman et al. 1995, Liseau 1999, \& Greaves
et al. 2000), perhaps because the angular size of the source as measured
in the infrared is significantly smaller than the beam of available
telescopes. Another possibility is that circumstellar molecular gas is not 
present because it is rapidly photodissociated under the intense stellar 
ultraviolet field of the central A0V star. Holweger, Hempel, \& Kamp (1999)
have detected two narrow, circumstellar \ion{Ca}{2} K absorption 
features, one component close to the stellar velocity and another weaker
one blueshifted by $\sim$10 km/sec which they attribute to possible
circumstellar gas.

\section{Observations}
HR 4796A was observed with the \emph{Far Ultraviolet Spectroscopic Explorer} 
(\emph{FUSE}) for 12.4 ksec on 2002 April 25 in histogram mode, using the 
low resolution 30$\arcsec$ $\times$ 30$\arcsec$ aperture which covers the 
wavelength range between 905 and 1187 \AA. This wavelength coverage allows us 
to search for circumstellar absorption in the ground states of \ion{C}{2} 
$\lambda$1036.3 and \ion{O}{1} $\lambda$1039.2, and in Lyman series H$_{2}$. 
The \emph{FUSE} satellite has four channels (LIF 1, SIC 1, LIF 2, SIC 2) 
which form two nearly identical ``sides'' (labeled 1 and 2) that each consist 
of a LIF grating, a SIC grating, and a detector. Each detector is divided 
into two independent segments (A and B), seperated by a small gap. The eight 
partially overlapping spectra that fall on different portions of the two 
detectors (LIF 1A, LIF 1B, etc.) cover the entire wavelength range. A 
description of the on orbit performance of \emph{FUSE} is described in Sahnow
et al. (2000) with a conservative estimate for the spectral resolution (R = 
$\lambda / \Delta \lambda$ = 15000) of the spectrograph. The data were 
calibrated at Johns Hopkins using the CALFUSE 1.8.7 pipeline.

HR 4796A was observed with Space Telescope Imaging Spectrograph (STIS) in
echelle mode on the \emph{Hubble Space Telescope} (\emph{HST}) for 930 sec 
and 550 sec on 2002 August 22, with the E140H and E230H gratings respectively
using the 0.2\arcsec$\times$0.09\arcsec slit. The spectra covered the 
wavelength regions 1425 \AA \ - 1620 \AA \ and 1880 \AA \ - 2155 \AA \ with a
spectral resolution (R = $\lambda / \Delta \lambda$ = 220,000). This 
wavelength coverage allows us to search for circumstellar absorption in the 
ground state of \ion{Zn}{2} $\lambda$2026.1 and in CO (A-X). The data were 
calibrated at the Space Telescope Science Institute using the standard
STIS software package \emph{calstis}, executed under the OPUS pipeline.

We observe absorption in the ground states of \ion{Si}{2} at 1526.14 \AA, 
\ion{Fe}{2} at 1608.45 \AA, and \ion{Zn}{2} at 2026.14 \AA \ along the line 
of sight to HR 4796A (see Figure 1). The \ion{Si}{2} and \ion{Fe}{2} lines 
possess two absorption components, one at about -5 km/sec and another at 
about -15 km/sec, consistent with the \ion{Ca}{2} results from Holweger et 
al. (1999). We believe that both gas components are interstellar. 
Circumstellar gas in Keplerian orbit at 70 AU from an A0V star would have a 
velocity of $\sim$5.5 km/sec. Therefore, any circumstellar gas features will 
be indistinguishable from interstellar lines on the basis of line width. 
However, circumstellar gas at 70 AU from an A0V should be radiatively pumped, 
populating the excited fine structure lines. We observe no absorption in the
\ion{Si}{2}$^{*}$ line at 1533.43 \AA, suggesting that the \ion{Si}{2} gas is
interstellar. Since the \ion{Fe}{2} line has the same velocity structure as 
the \ion{Si}{2} line, this material is probably also interstellar. Even 
though the \ion{Zn}{2} line only possesses one velocity component, its 
velocity is similar to that of the \ion{Si}{2} and \ion{Fe}{2} components 
near the stellar velocity. 

We observe no absorption in the H$_{2}$ Lyman (5,0) band or the CO A-X (1,0) 
band toward this star. The 3$\sigma$ upper limits on the H$_{2}$ 
line-of-sight column densities in the J = 0 and J = 1 rotational levels are 
measured, from the P(1) and R(0) transitions at 1036.54 \AA \ and 
1038.16 \AA, to be $N(J = 0) \leq 10^{15}$ cm$^{-2}$ and 
$N(J = 1) \leq 3.7 \times 10^{15}$ cm$^{-2}$ using wavelengths and oscillator 
strengths from Abgrall et al. (1993). The 3$\sigma$ upper limits on the CO 
line-of-sight column densities in the J = 1 and J = 2 rotational levels are 
measured, from the R(0) and R(1) transitions at 1544.45 \AA \ and 1544.39.73
\AA, to be $N(J = 1) \leq 1.2 \times 10^{13}$ cm$^{-2}$ and $N(J = 2) \leq 
2.4 \times 10^{13}$ cm$^{-2}$ using wavelengths and oscillator strengths from
Morton \& Noreau (1994). Similarly, we observe no absorption in \ion{C}{2} 
$\lambda$1036.3 and \ion{O}{1} $\lambda$1039.2, and no circumstellar 
absorption in \ion{Zn}{2} $\lambda$2026.1 a tracer for atomic hydrogen in the
interstellar medium (York \& Jura 1982). The 3$\sigma$ upper limits on these 
column densities are 
$N($\ion{C}{2}$) \leq 1.4 \times 10^{14}$ cm$^{-2}$, 
$N($\ion{O}{1}$) \leq 2.8 \times 10^{15}$ cm$^{-2}$, and
$N($\ion{Zn}{2}$) \leq 2.6 \times 10^{12}$ cm$^{-2}$ and are estimated using 
oscillator strengths from Morton (1991). We calculate equivalent width upper 
limits assuming circumstellar line widths of 5.5 km/sec for the STIS data.
We calculate equivalent width upper limits assuming circumstellar line
widths of 2 resolution elements, after binning the data to a resolution of 
0.05 \AA \ or $\sim$29 km/sec for the \emph{FUSE} data.

\section{Disk Vertical Structure}
Circumstellar gas in the low density environments around Vega-type stars is 
primarily heated via photoelectrons ejected from dust grains by stellar 
ultraviolet photons and primarily cooled by radiation from the [\ion{C}{2}] 
fine structure line at 157.5 $\mu$m (Kamp \& van Zadelhoff 2001). 
HR 4796A possesses an M-type companion, located 7.7$\arcsec$ away from the 
star. The companion has a \emph{ROSAT} 0.1 - 2.4 keV luminosity, L$_{x}$ = 
1.5 $\times$10$^{29}$ ergs/sec (Jura et al. 1998) which corresponds to an 
X-ray flux of $F_{x}$ $\leq$ 2 $\times$ 10$^{-4}$ ergs s$^{-1}$ cm$^{-2}$ at 
the position of HR 4796A. We estimate the photoelectric and X-ray heating 
rates for the gas around HR 4796A to determine whether X-ray heating plays an 
important role in this circumstellar environment. 

The photoelectric heating rate for micron-sized silicate grains, like the 
ones found around HR 4796A (Augereau et al. 1999; Li \& Lunine 2003), is
\begin{equation}
\Gamma_{pe} = 2.5 \times 10^{-4} \sigma \epsilon \chi n_{H}
\end{equation}
(Kamp \& van Zadelhoff 2001) where $n_{H}$ is the number density of hydrogen,
$\sigma$ is the grain cross section per hydrogen nucleus, $\epsilon$ is the 
photoelectric efficiency, and $\chi$ is the 912 - 1110 \AA \ photon flux 
measured in units of the Habing field ($F_{H}$ = 1.2 $\times$ 10$^{7}$ 
cm$^{-2}$ s$^{-1}$). At a distance of 70 AU from HR 4796A, we estimate $\chi$
= 460 from our \emph{FUSE} spectra, roughly consistent with a Kurucz stellar 
atmosphere with $T_{eff}$ = 10,000 K, $\log g$ = 4.0, and solar metallicity 
(Fajardo-Acosta, Telesco, \& Knacke 1998). For large silicate grains, the 
dust UV cross section $\sigma$ = 2.34 $\times$ 10$^{-21}$/$\delta$ cm$^{2}$ 
H-atom$^{-1}$ where $\delta$ is the gas:dust ratio and $\epsilon$ = 0.06 for 
low temperatures. If the gas:dust ratio is 4 ($\sigma$ = 5.85 $\times$ 
10$^{-22}$ cm$^{2}$ H-atom$^{-1}$), $\epsilon$ = 0.06, and $\chi$ = 460, then
the photoelectric heating rate is $\Gamma_{pe}$ = 4.0 $\times$ 10$^{-24}$ 
$n_{H}$ ergs cm$^{-3}$ s$^{-1}$. 

The X-ray heating rate for a cold neutral gas is
\begin{equation}
\Gamma_{x} = f_{h} \sigma_{x} F_{x} n_{H}
\end{equation}
where $f_{h} (\sim0.3)$ is the fraction of the absorbed X-ray energy that 
heats the gas, $\sigma_{x}$ is the X-ray cross section due to all atoms per 
hydrogen nucleus, and $F_{x}$ is the X-ray flux. If the bulk of the X-ray 
energy is carried by 0.5 keV photons, then the cross section $\sigma_{x}$ = 
1.65 $\times$ 10$^{-21}$ cm$^{2}$ (Maloney, Hollenbach, \& Tielens 1996). If 
$F_{x}$ $\leq$ 2 $\times$ 10$^{-4}$ ergs s$^{-1}$ and $\sigma_{x}$ = 1.65 
$\times$ 10$^{-21}$ cm$^{2}$, then $\Gamma_{x}$ $\leq$ 9.9 $\times$ 
10$^{-26}$ $n_{H}$ ergs cm$^{-3}$ s$^{-1}$, suggesting that X-ray heating is 
not significant in the optically thin regime. X-ray heating in more gaseous 
circumstellar disks is even less important because the disk becomes optically
thick to the X-ray radiation.

Kamp \& van Zadelhoff have derived an analytic approximation for the gas 
temperature assuming that the gas is heated by photoelectrons ejected by 
dust grains and is cooled via [\ion{C}{2}] line emission at 157.7 $\mu$m. 
In this case, the gas temperature 
\begin{equation}
T_{gas} = \frac{91.98 \ \textup{K}}
{\log [2(\frac{\epsilon_{C} A_{10} h \nu_{10}}
              {2.5 \times 10^{-4} \epsilon \sigma \chi} 
       - 1)]}
\end{equation}
where $\epsilon_{C}$ is the carbon abundance, $A_{10}$ and $h \nu_{10}$ are 
the Einstein A coefficient and energy carried by the photon in the \ion{C}{2}
transition. For the 157.7 $\mu$m [\ion{C}{2}] line, $A_{10}$ = 2.4 $\times$
10$^{-6}$ s$^{-1}$ and $h \nu_{10}$ = 1.27 $\times$ 10$^{-14}$ ergs. If
the circumstellar environment around HR 4796A possesses a solar
carbon abundance, then $\epsilon_{C}$ = 4.6 $\times$ 10$^{-4}$ and the gas
temperature T$_{gas}$ = 65 K. If the carbon abundance is interstellar
$\epsilon_{C}$ = 1.4 $\times$ 10$^{-4}$, then the cooling via carbon emission
will be less efficient and the circumstellar gas will be hotter. In Figure 2,
we plot the gas temperature as a function of the circumstellar gas mass (or 
gas:dust ratio) as inferred from equation (3). For HR 4796A, this 
approximation only holds for the outer portions of the disk. For example, 
\ion{O}{1} is an important coolant at the inner edge of the disk midplane. In
addition, this approximation breaks down for HR 4796A models with gas:dust 
ratio larger than a few. In higher gas:dust ratio models, the implied gas 
densities are large enough to allow CO to become self shielding, unlike the 
circumstellar disk models around Vega, because the HR 4796A circumstellar 
disk is more compact. In these cases, the cooling also becomes more 
complicated. 

We assume that the circumstellar gas is in hydrostatic equilibrium. In
this model, the gas number density, $n$, can be written as a function of the 
height above the disk midplane, $z$,
\begin{equation}
n(z)=n_{o}e^{-\frac{1}{2}(\frac{z}{H})^2} \ \textup{where} 
\ \frac{H}{R}=(\frac{kT_{gas}R}{\mu GM_{*}})^\frac{1}{2}
\end{equation}
where $H$ is the disk thickness parameter, $R$ is the distance to the star, 
$M_{*}$ is the stellar mass, and $\mu$ is the mean molecular weight of the 
gas. If the gas is primarily H$_{2}$ with [He]/[H$_{2}$] = 0.2 so that $\mu$
= 3.9 $\times$ 10$^{-24}$ g. If the gas is primarily \ion{H}{1} with
[He]/[H$_{2}$] = 0.1, then $\mu$ = 2.2 $\times$ 10$^{-24}$ g. Since HR 4796A 
is not viewed edge-on, successful detection of circumstellar gas (or the 
establishment of meaningful gas mass upper limits) via absorption line 
studies requires that the disk thickness parameter, $H$ $\sim$ $R \sin i$. 
Smaller thickness parameters lead to less puffed-up disks which provide fewer
gas atoms and molecules along the line of sight. The inclination and the 
radius of the HR 4796A disk has been measured to be $\sim$17$\arcdeg$ from 
edge-on and 70 AU (Schneider et al. 1999) suggesting that a disk thickness 
parameter of $\sim$10 AU is necessary. If the gas is molecular, $M_{*}$ = 2.4
$M_{\sun}$, and $T_{gas}$ = 65 K, then the thickness parameter $H$ = 6.1 AU 
at $R$ = 70 AU and our line of sight intercepts the gas at 3.4 $H$ above the 
disk. If the gas is atomic, then the thickness parameter $H$ = 8.1 AU at $R$ 
= 70 AU and our line of sight intercepts the gas at 2.5 $H$ above the disk.

\section{Gas Mass Estimates}
We estimate the gas mass upper limits from our line-of-sight column density 
upper limits for molecular and atomic gases assuming that the gas species are
well-mixed and in hydrostatic equilibrium. (i.e. If the gas is molecular, we 
will assume a disk thickness parameter $H$ = 6.1 AU for both CO and H$_{2}$ 
despite the fact that CO molecules are more massive than $H_{2}$ molecules. 
If the gas is atomic, we will assume a disk thickness parameter $H$ = 8.1 AU 
for \ion{C}{2}, \ion{O}{1}, and \ion{Zn}{2}.) To estimate the circumstellar 
gas mass from the column density upper limits, we must assume a particular 
gas mass surface density, $\Sigma$. For simplicity, we assume that the gas 
and dust have the same radial distribution. If the dust and gas are not 
cospatial then photoelectric heating will not effectively warm the gas and 
the gas temperature will be significantly colder than 65 K.
\begin{eqnarray}
\Sigma(R) \: & = &\Sigma_0 
\: \left(
\begin{array}{ccc}
d \left(\frac{R}{R_0}\right)^{-2.5}
& \textup{if} & R < R_1\\
e^{-\frac{(R-R_0)^2}{\delta R_0^2}}
& \textup{if} & R_1 \leq R \leq R_0\\
\left(\frac{R}{R_0}\right)^{-2.5}& \textup{if} & R > R_0
\end{array} \; \; \right)\;
\end{eqnarray}
(Klahr \& Lin 2000) where $R_{0}$ = 70 AU is the peak of the dust 
distribution, $\delta R_{0}$ = 10 AU, and $d$ = 0.9 (which corresponds to 
$R_{1}$ = 68.1 AU). We can write an expression for the gas density from
equations (4) and (5)
\begin{equation}
n(r,z) = \frac{\Sigma(r)}{\sqrt{2 \pi} H(r)} e^{-\frac{1}{2}(\frac{z}{H(r)})^2}
\end{equation}
We estimate the column density by integrating the mass density in equation 
(6) along the line-of-sight. If the line of sight is inclined $i \arcdeg$ 
from the disk midplane, then the estimated column density, $N$, is
\begin{equation}
N = \frac{1}{m_{s} \cos i \sqrt{2 \pi}} 
    \int_{R_{in}}^{R_{out}} \frac{\Sigma(R)}{H(R)} 
    e^\frac{- \sin^{2} i R^2}{2 H(R)^2} dR
\end{equation}
where $m_{s}$ is the mass of the species of interest, $i$ = 17$\arcdeg$, 
$T_{gas}$ = 65 K, $M_{*}$ = 2.4 $M_{\sun}$, $R_{in}$ = 60 AU and $R_{out}$ = 
80 AU. 

If we set the estimated column density to be less than or equal to the
observed column density upper limit, we can establish an upper limit for 
$\Sigma_{0}$. Once $\Sigma_{0}$ is known, estimating the circumstellar gas 
mass upper limit only requires integrating the mass surface density. For 
molecular gas around HR 4796A, the gas mass in species $s$ can be written as 
the following expression:
\begin{equation}
M_{gas} \leq 2.9 \times 10^{-4} \left( \frac{N}{10^{15} \textup{cm}^{-2}} 
    \right) \left( \frac{m_{s}}{\mu} \right) M_{\earth}
\end{equation}
where $N$ is the measured column density. The J = 0 and J = 1 H$_{2}$ column 
density upper limits listed in Table 3 suggest H$_{2}$ gas masses of 
$\leq$2.5$\times$10$^{-4}$ $M_{\earth}$ and $\leq$9.1$\times$10$^{-4}$ 
$M_{\earth}$, respectively. This result suggests that the HR 4796A 
circumstellar disk possesses very little molecular gas.

The nondetection of H$_{2}$ around HR 4796A may not be surprising because 
HR 4796A, an A0V star, produces a high ultraviolet photon flux which is 
capable of rapidly photodissociating molecular gas. Therefore, we also 
searched for absorption due to circumstellar \ion{O}{1}, \ion{C}{2}, and 
\ion{Zn}{2}, a tracer for atomic hydrogen in the interstellar medium
(York \& Jura 1983). We can estimate the gas mass assuming that the bulk of 
the gas is predominately atomic.
\begin{equation}
M_{gas} \leq 3.3 \times 10^{-5} \left( \frac{N}{10^{15} \textup{cm}^{-2}} 
    \right) \left( \frac{m_{s}}{\mu} \right) M_{\earth}
\end{equation}
We list gas mass upper limits for each of the atomic species in Table 3 using
equation (9). We also infer atomic hydrogen disk mass upper limits assuming 
that the composition of the circumstellar gas is approximately solar. 
\begin{equation}
M_{H} \leq 1.4 \times 10^{-8} \frac{1}{\epsilon_{s}} 
\left( \frac{N}{10^{12} \textup{cm}^{-2}} \right)
M_{\earth}
\end{equation}
where $\epsilon_{s}$ is the abundance of the species of interest. 
For $\epsilon_{O}$ = 8.1 $\times$ 10$^{-4}$, $\epsilon_{C}$ = 4.6 $\times$ 
10$^{-4}$, and $\epsilon_{Zn}$ = 3.8 $\times$ 10$^{-8}$ (Gray 1992), we find 
$M_{H}$ $\leq$1.0 $M_{\earth}$. Since submillimeter continuum measurements 
estimate $\geq$0.25 $M_{\earth}$ dust around HR 4796A (Greaves et al. 2000), 
our hydrogen gas mass upper limit corresponds to a gas:dust mass ratio of 
$\leq$4.0.

If circumstellar gas is in hydrostatic equilibrium, then any gas mass 
estimates are dependent on the inferred gas temperature. The density depends
on the exponential of -(1/H)$^2$; therefore, small changes in the disk
thickness parameter can lead to dramatic changes in the inferred gas mass.
We plot the inferred \ion{H}{1} mass from our \ion{Zn}{2} column density
upper limit as a function of gas temperature in Figure 3. For example,
changing the gas temperature from 65 K to 28 K corresponds to changing
the disk thickness parameter from 8.1 AU to 5.3 AU and the inferred \ion{H}{1}
mass from $\leq$1.0 $M_{\earth}$ to 40 $M_{\earth}$. This corresponds to
a gas:dust ratio $\leq$160, approximately the interstellar gas:dust ratio.

\section{The Gas Temperature and Disk Chemistry}
Kamp \& van Zadelhoff (2001) have produced detailed numerical models
of circumstellar gas heating and cooling around main sequence A-type stars to 
calculate the gas temperature and chemistry. Given a prescribed gas and dust 
density distribution, these models solve the chemistry and energy balance of 
the gas self-consistently on a two dimensional grid. The computation proceeds
along a number of rays originating at the star's position under various 
angles. Typically, 500 radial gridpoints and 30 angles covering about 3
scaleheights are considered. The heating in low density disks, like the one 
found around HR 4796A, is dominated by photoelectrons ejected from 
circumstellar dust grains by stellar ultraviolet photons. Kamp \& van 
Zadelhoff (2001) use ATLAS9 models to estimate stellar photospheres and 
assume that circumstellar dust grains are large silicate grains with radius, 
$a$ = 3 $\mu$m, and density, $\rho$ = 3.0 g cm$^{-3}$. They assume a dust UV 
extinction cross section $\sigma_{UV}$ = 2.34 $\times$ 10$^{-21}/\delta$ 
cm$^{2}$ (H-atom)$^{-1}$ and a fixed fraction of vibrationally excited 
H$_{2}$, $f_{H_{2}^{*}}$ = 1.0 $\times$10$^{-5}$. 

We have improved the Kamp \& van Zadelhoff (2001) models by making the 
following changes: (1) including cosmic ray reactions, (2) lowering the 
temperature for CO freeze out to $\sim$20 K, (3) including an escape 
probability formalism for line photons (Tielens \& Hollenbach 1985; TH85), 
and (4) calculating the statistical equilibrium for \ion{C}{2} following
the approach used for \ion{O}{1} and CO by Kamp \& van Zadelhoff (2001). 
Cosmic ray reactions are added to make the code more general. This change 
does not affect the result described here because the chemistry around main 
sequence A-type stars is driven by stellar photons. The escape probability 
formalism is necessary because the main cooling lines, such as [\ion{O}{1}] 
and CO, become optically thick for disks with gas:dust ratios of 10 and 100. 
The escape probability of a line photon, $\beta$, is conservatively estimated 
from the line optical depth towards the star at each gridpoint. Since the 
optical depth perpendicular to the disk is much lower than the optical depth
in the disk midplane, our estimates yield reasonable lower limits for the line
cooling and therefore reasonable upper limits for the gas temperature. We 
also include a model which does not contain the one dimensional escape 
probability approximation to estimate the possible error arising from this 
approach.

We have modeled the circumstellar gas around HR 4796A using the model
described above with a gas density distribution,
\begin{equation}
n(r) = n_{i} \left( \frac{R}{R_{i}} \right)^{-2.5} 
e^{-\frac{1}{2}(\frac{z}{H})^2}
\end{equation}
where $R_{i}$ is the inner disk radius and $n_{i}$ is the gas density
at $R_{i}$. We assume that the density distribution is fixed a priori
with a disk thickness parameter $H \propto R$ and $H/R = 0.12$. For HR 4796A, 
we further assume $R_{i}$ = 60 AU and an outer disk radius $R_{o}$ = 80 AU. 
The disk thickness parameter  $H = 8.4$ at 70 AU is approximately consistent 
with our estimate for an atomic gas scale height, $H$ = 8.1 AU (see Section 
4). We fix the dust mass $M_{dust}$ = 0.25 $M_{\earth}$ and calculate the gas
chemistry and temperature distribution assuming gas:dust ratios of 2, 10, and
100. An additional gas:dust ratio = 100 model is calculated assuming a 
constant escape probability, $\beta$=1. The model parameters are summarized 
in Table 4.

The calculated temperature structures for the HR 4796A circumstellar disk 
models are shown in Figure 3. Deep inside the disk, [\ion{O}{1}] and 
[\ion{C}{2}] cooling balance photoelectric heating and heating by H$_2$ 
formation. With increasing gas mass, fine-structure line cooling becomes more
efficient while photoelectric heating remains constant. The photoelectric 
heating rate does not change because the dust mass is fixed in all of the 
models. Thus, more massive disk models possess significantly cooler top 
layers. Since the one dimensional escape probability is only a crude 
approximation to the actual escape probabilty, we also model the HR 4796A 
disk with $\beta =1$ for a gas:dust ratio of 100 (in which the [\ion{O}{1}] 
63 $\mu$m line has an optical depth of $\sim$10 at 80 AU). In addition to 
escape along the line of sight to the star, line radiation can also escape 
perpendicular to the disk and through the outer radius. Therefore, models 
which include the one dimensional escape probability approximation yield an 
upper limit to the disk midplane gas temperature while models with $\beta =1$
yield a lower limit. The gas temperatures estimated with and without the one 
dimensional escape probability approximation differ by less than a factor of 
2 or by less than 30\% in the disk thickness parameter.

In Table 5, we list the predicted gas temperatures averaged over one
scaleheight at 70 AU and the predicted \ion{H}{1}, H$_2$, \ion{Zn}{2},
\ion{C}{2} and CO gas column densities assuming a line-of-sight inclined 
17$\arcdeg$ from the disk midplane. The column densities estimated from
these detailed heating and cooling models suggest that the measured
column densities are not constraining, that the gas:dust ratio = 100
model is allowed for all species, except for \ion{C}{2}. The measured
\ion{C}{2} column density, N(\ion{C}{2}) = $1.4 \times 10^{14}$ cm$^{-2}$,
is most consistent with the $M_{gas}$ = 0.5 $M_{\earth}$ model calculated 
using the 1-D escape probability approximation which predicts N(\ion{C}{2}) = 
$1.5 \times 10^{14}$ cm$^{-2}$. This result is consistent with the simple
analytical estimate for the gas mass, $M_{gas}$ $\leq$ 1.0 $M_{\earth}$

\section{Discussion}
In general, the simple analytic model and the detailed chemical model 
estimate similar temperatures and gas masses for the circumstellar disk 
around HR 4796A. The discrepancy between the predicted \ion{Zn}{2} column 
densities from the simple analytical model and the detailed chemical model 
may be due to (1) the lack of detailed structure in the simple model and (2) 
the different surface density distributions used. In the simple model, the 
entire disk is assumed to be either atomic or molecular even though the disk 
possesses a more complicated chemical structure. For the low gas:dust ratio 
case, the upper regions of the disk are predominantly atomic while the 
midplane consists of cooler molecular material. The line-of-sight intercepts 
the upper layer of atomic hydrogen but few molecules of cooler H$_{2}$. 
The surface density, $\Sigma$, in the simple model is a Gaussian superimposed
on a $R^{-2.5}$ power law while the surface density in the chemical model is 
only described by a $R^{-2.5}$ power law. Despite the difference in assumed 
functions, the surface densities in the models agree to within a factor of 
two at all radii in the disk suggesting that this difference can not account 
for the discrepancy in the predicted \ion{Zn}{2} column densities.

The circumstellar disk around HR 4796 probably possesses $\leq$1.0 
$M_{\earth}$ \ion{H}{1} which corresponds to a gas:dust ratio $\leq$4.0
although a gas:dust ratio of 100 can not be definitively ruled out. The 
atmosphere of a giant planet contains approximately 275 $M_{\earth}$ of 
H$_{2}$ (Tholen, Tejfel, \& Cox 2000). This is substantially more than either 
$\leq$1.0 $M_{\earth}$ circumstellar gas we infer to exist around HR 4796A or
even the $\leq$5.0 $M_{\earth}$ gas that would exist if the gas:dust ratio 
were interstellar. Therefore, there is currently not enough circumstellar gas 
around HR 4796A to form the atmosphere of a Jovian planet. If Jupiters exist 
around HR 4796A, they must have formed within $\sim$8 Myr, significantly less
time than is required by multi-stage accumulation (10 - 100 Myr).

\section{CONCLUSIONS}
We have obtained ultraviolet spectra (905 \AA \ - 1187 \AA, 1425 \AA \ - 
1620 \AA, 1880 \AA \ - 2155 \AA) of the nearby, debris disk object HR 4796A 
using \emph{FUSE} and STIS on \emph{HST}. We argue the following:

1. The X-ray flux from the M-dwarf companion HR 4796B does not significantly 
heat circumstellar gas around HR 4796A. The estimated X-ray heating rate is 
at least $\sim$40 times smaller than the estimated photoelectric heating rate.

2. Circumstellar gas around HR 4796A may be warmed to a temperature $T_{gas}$ 
$\sim$ 65 K by photoelectric heating if the gas:dust ratio is 4, suggesting
a molecular gas disk thickness parameter $H$ = 6.1 AU and an atomic gas disk 
thickness parameter $H$ = 8.1 AU at 70 AU from the star. This is 
approximately consistent with detailed heating and cooling models which 
suggest a maximum gas temperature, $T_{gas}$ 61 K, for a gas:dust ratio of 2,
assuming a 1-D approximation for the escape probability of line photons.

3. The circumstellar disk around HR 4796A possesses very little $H_{2}$.
We estimate that $\leq$1.1 $\times$ $10^{-3}$ $M_{\earth}$ H$_{2}$ exists 
around HR 4796A assuming that gas is in hydrostatic equilibrium, is 
well-mixed, and has a temperature T$_{gas}$ = 65 K.

4. We estimate that $\leq$1.0 $M_{\earth}$ atomic hydrogen exists around 
HR 4796A. This corresponds to a gas:dust mass ratio $\leq$4.0 assuming that 
gas is in hydrostatic equilibrium, is well-mixed, and has a temperature 
T$_{gas}$ = 65 K. 

5. Measurements of \ion{C}{2} column density may provide the strongest
constraint on the gas mass in the HR 4796A circumstellar disk. While the
measured H$_{2}$, CO, and \ion{Zn}{2} column densities are consistent
with the column densities predicted for the gas:dust ratio = 100 model,
the measured \ion{C}{2} column denisty N(\ion{C}{2}) = $1.4 \times 10^{14}$
cm$^{-2}$ is most consistent with the gas:dust ratio = 2 model.

6. The circumstellar disk around HR 4796A is too gas depleted to support
the formation of giant planets. Any Jovian planets in this system must
have formed on a timescale $\leq$8 Myr.

\acknowledgements
We would like to thank G. Bryden, M. Jura, M. Werner, K. Willacy, and
our anonymous referee for helpful comments and suggestions. This research has
been supported by funding from the National Aeronautics and Space 
Administration (NASA) and was partially carried out at the Jet Propulsion 
Laboratory which is managed by the California Institute of Technology under a
contract with NASA.

\begin{deluxetable}{lll}
\singlespace
\tablecaption{HR 4796A Properties} 
\tablehead{
    \colhead{Quantity} &
    \colhead{Adopted Value} &
    \colhead{Reference} \\
}
\tablewidth{0pt}
\tablecolumns{3}
\startdata
    Primary Spectral Type & A0V & 1 \\
    Distance & 67.1$\pm$3.4 pc & 2 \\
    Effective Temperature ($T_{eff}$) & 10,000$\pm$500 K & \\
    Stellar Radius ($R_{*}$) & 1.7 $R_{\sun}$ & 2 \\
    Stellar Luminosity ($L_{*}$) & 21 $L_{\sun}$ & 2 \\
    Stellar Mass ($M_{*}$) & 2.4 $M_{\sun}$ & \\
    Rotational Velocity ($v\sin i$) & 152$\pm$8 km/sec & 3 \\
    Fractional Dust Luminosity & & \\
        \ \ \ \ \ ($L_{IR}/L_{*}$) & 5$\times$10$^{-3}$ & 2 \\ 
    Estimated Age & 8$\pm$2 Myr & 4 \\
    Grain Temperature & 110 K & 2 \\
    Dust Distance ($R_{0}$) & 70$\pm$1 AU & 5 \\
    Dust Mass ($M_{dust}$) & $\geq$0.25 $M_{\earth}$ & 6 \\
\enddata
\tablerefs{(1) Hoffleit \& Warren (1991);
           (2) Jura et al. (1998);
           (3) Royer et al. (2002);
           (4) Stauffer et al. (1995);
           (5) Schneider et al. (1999);
           (6) Greaves et al. (2000)
          }
\end{deluxetable}

\begin{deluxetable}{llcccc}
\singlespace
\tablecaption{Interstellar Gas Properties} 
\tablehead{
    \omit &
    \omit &
    \colhead{Ground State} &
    \omit & 
    \omit &
    \omit \\
    \colhead{Species} &
    \colhead{Wavelength} &
    \colhead{Energy} &
    \colhead{$W_{\lambda}$} & 
    \colhead{N} &
    \colhead{$v$} \\
    \omit &
    \colhead{(\AA)} &
    \colhead{(cm$^{-1}$)} &
    \colhead{(m\AA)} & 
    \colhead{(cm$^{-2}$)} &
    \colhead{(km/sec)} \\
}
\tablewidth{0pt}
\tablecolumns{3}
\startdata
  \ion{Si}{2} & 1526.14 & 0.00 & 65 & $\geq$2.4$\times$10$^{13}$ & -5.0 \\
  \ion{Si}{2} & 1526.14 & 0.00 & 20 & $\geq$7.4$\times$10$^{12}$ & -14.4 \\
  \ion{Fe}{2} & 1608.45 & 0.00 & 14 & $\geq$9.9$\times$10$^{12}$ & -5.5 \\
  \ion{Fe}{2} & 1608.45 & 0.00 & 15 & $\geq$1.1$\times$10$^{13}$ & -14.6 \\
  \ion{Zn}{2} & 2026.14 & 0.00 &  7 & $\geq$3.8$\times$10$^{11}$ & -6.0 \\
\enddata
\end{deluxetable}

\begin{deluxetable}{lccccccc}
\singlespace
\tablecaption{Circumstellar Gas Upper Limits} 
\tablehead{
    \omit & 
    \omit & 
    \colhead{Ground State} &
    \omit & 
    \omit & 
    \omit & 
    \omit & 
    \colhead{Gas:Dust} \\
    \colhead{Species} &
    \colhead{Wavelength} & 
    \colhead{Energy} & 
    \colhead{$W_{\lambda}$} & 
    \colhead{N} &
    \colhead{$M_{gas}$} &
    \colhead{$M_{H}$} &
    \colhead{Ratio} \\
    \omit & 
    \colhead{(\AA)} & 
    \colhead{(cm$^{-1}$)} & 
    \colhead{(m\AA)} & 
    \colhead{(cm$^{-2}$)} &
    \colhead{($M_{\earth}$)} &
    \colhead{($M_{\earth}$)} &
    \omit \\
}
\tablewidth{0pt}
\tablecolumns{8}
\startdata
    \ion{C}{2} & 1036.33 & 0.00 & $\leq$160 & $\leq$1.4$\times$10$^{14}$ &
       $\leq$2.4$\times$10$^{-5}$ & $\leq$4.5$\times$10$^{-3}$ & 
       $\leq$0.018 \\
    \ion{O}{1} & 1039.23 & 0.00 & $\leq$250 & $\leq$2.8$\times$10$^{15}$ &
       $\leq$6.4$\times$10$^{-4}$ & $\leq$0.051 & $\leq$0.20 \\
    \ion{Zn}{2} & 2026.14 & 0.00 & $\leq$5 & $\leq$2.6$\times$10$^{12}$ &
       $\leq$2.4$\times$10$^{-6}$ & $\leq$1.0 & $\leq$4.0 \\
    H$_{2}$ (J=0) & 1036.54 & 0.00 & $\leq$260 & $\leq$1.0$\times$10$^{15}$ & 
       $\leq$2.5$\times$10$^{-4}$ & $\leq$2.5$\times$10$^{-4}$ & 
       $\leq$1.0$\times$10$^{-3}$ \\
    H$_{2}$ (J=1) & 1038.16 & 118.16 & $\leq$300 & $\leq$3.7$\times$10$^{15}$ & 
       $\leq$9.1$\times$10$^{-4}$ & $\leq$9.1$\times$10$^{-4}$ & 
       $\leq$3.6$\times$10$^{-3}$ \\
    CO (J=1) & 1544.45 & 3.85 & $\leq$4 & $\leq$1.2$\times$10$^{13}$ &
       $\leq$4.1$\times$10$^{-5}$ & N/A & N/A \\
    CO (J=2) & 1544.39 & 11.58 & $\leq$4 & $\leq$2.4$\times$10$^{13}$ &
       $\leq$8.3$\times$10$^{-8}$ & N/A & N/A \\
\enddata
\end{deluxetable}

\begin{deluxetable}{lcccc}
\singlespace
\tablecaption{HR 4796A Model Parameters}
\tablehead{
    \colhead{Model} & 
    \colhead{M$_{\rm gas}$} & 
    \colhead{Gas:Dust} & 
    \colhead{$\sigma$} & 
    \colhead{escape approach} \\ 
    \omit & 
    \colhead{(M$_{\earth}$)} & 
    \colhead{Ratio} &
    \colhead{(cm$^2$/H-atom)} & 
    \omit \\ 
}
\tablewidth{0pt}
\tablecolumns{5}
\startdata
Model 1 & 0.5 & 2 & $1.17 \,10^{-21}$ & TH85 \\
Model 2 & 2.5 & 10 & $2.34\,10^{-22}$ & TH85 \\
Model 3 & 25.0 & 100 & $2.34\,10^{-23}$ & TH85 \\
Model 4 & 25.0 & 100 & $2.34\,10^{-23}$ & $\beta =1$ \\
\enddata
\end{deluxetable}

\begin{deluxetable}{lcccccc}
\singlespace
\tablecaption{Model Predicted Gas Temperatures and Column Densities} 
\tablehead{
  \omit & \colhead{Averaged} & \colhead{Predicted} & \colhead{Predicted} & 
    \colhead{Predicted} & \colhead{Predicted} & \colhead{Predicted} \\
  \colhead{Model} & \colhead{$T_{gas}$(70 AU)} & \colhead{$N_{\rm H}$} & 
    \colhead{$N_{\rm H2}$} & \colhead{$N_{\rm Zn+}$} & \colhead{$N_{\rm C+}$} &
     \colhead{$N_{\rm CO}$} \\
  \omit & \colhead{(K)} & \colhead{(cm$^{-2}$)} & \colhead{(cm$^{-2}$)} & 
    \colhead{(cm$^{-2}$)} & \colhead{(cm$^{-2}$)} & \colhead{(cm$^{-2}$)} \\
}
\tablewidth{0pt}
\tablecolumns{4}
\startdata
  Model 1 & 61 & $1.1 \times 10^{18}$ & $2.3 \times 10^{13}$ & 
    $4.2 \times 10^{10}$ & $1.5 \times 10^{14}$ & $4.2 \times 10^{6}$\\
  Model 2 & 45 & $5.3 \times 10^{18}$ & $8.4 \times 10^{13}$ & 
    $2.0 \times 10^{11}$ & $7.5 \times 10^{14}$ & $1.1 \times 10^{8}$\\    
  Model 3 & 52 & $5.3 \times 10^{19}$ & $1.0 \times 10^{15}$ & 
    $2.0 \times 10^{12}$ & $7.3 \times 10^{15}$ & $9.9 \times 10^{9}$\\
  Model 4 & 22 & & & & & \\
\enddata
\end{deluxetable}

\begin{figure}[ht]
\figurenum{1}
\plottwo{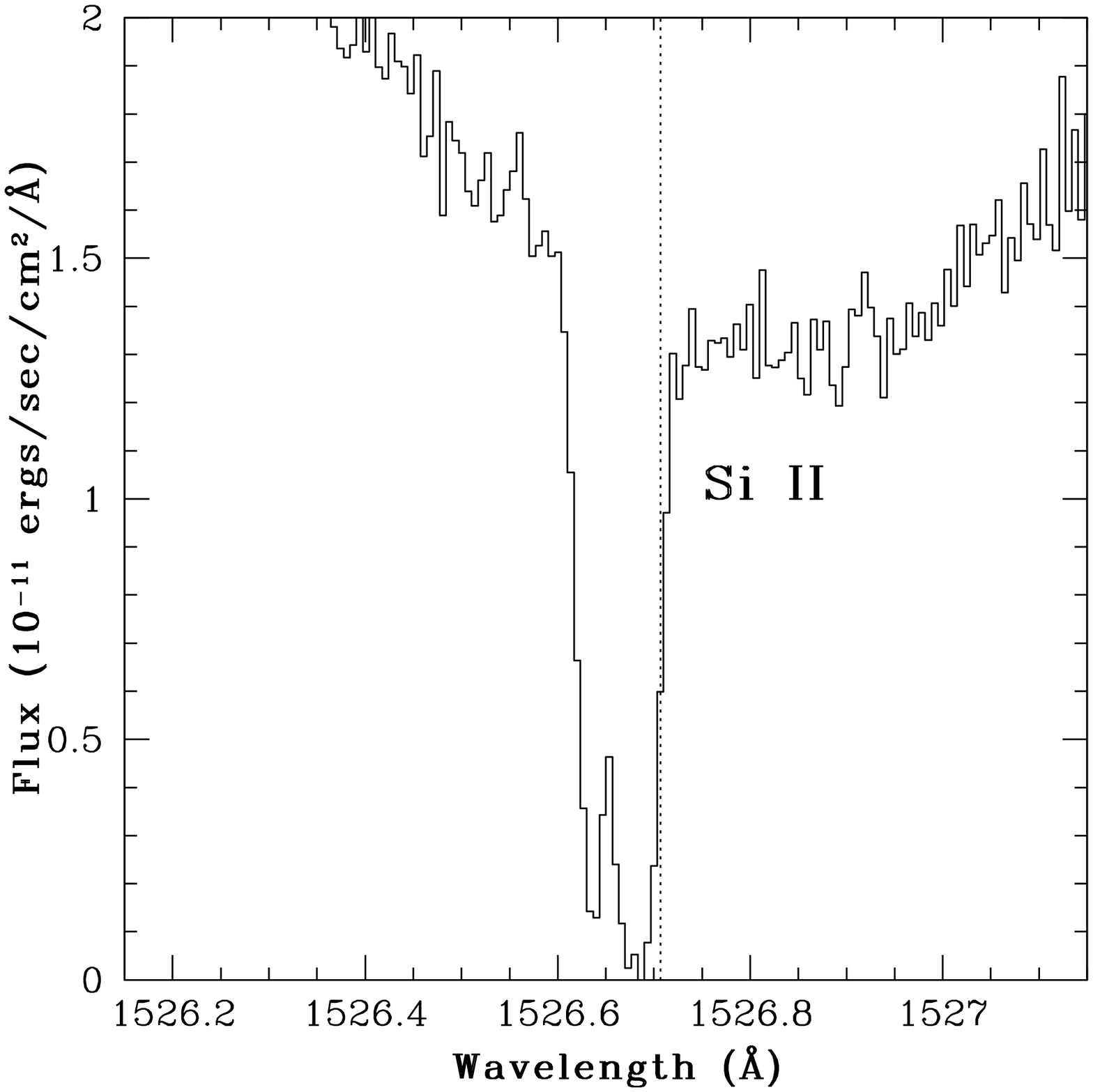}{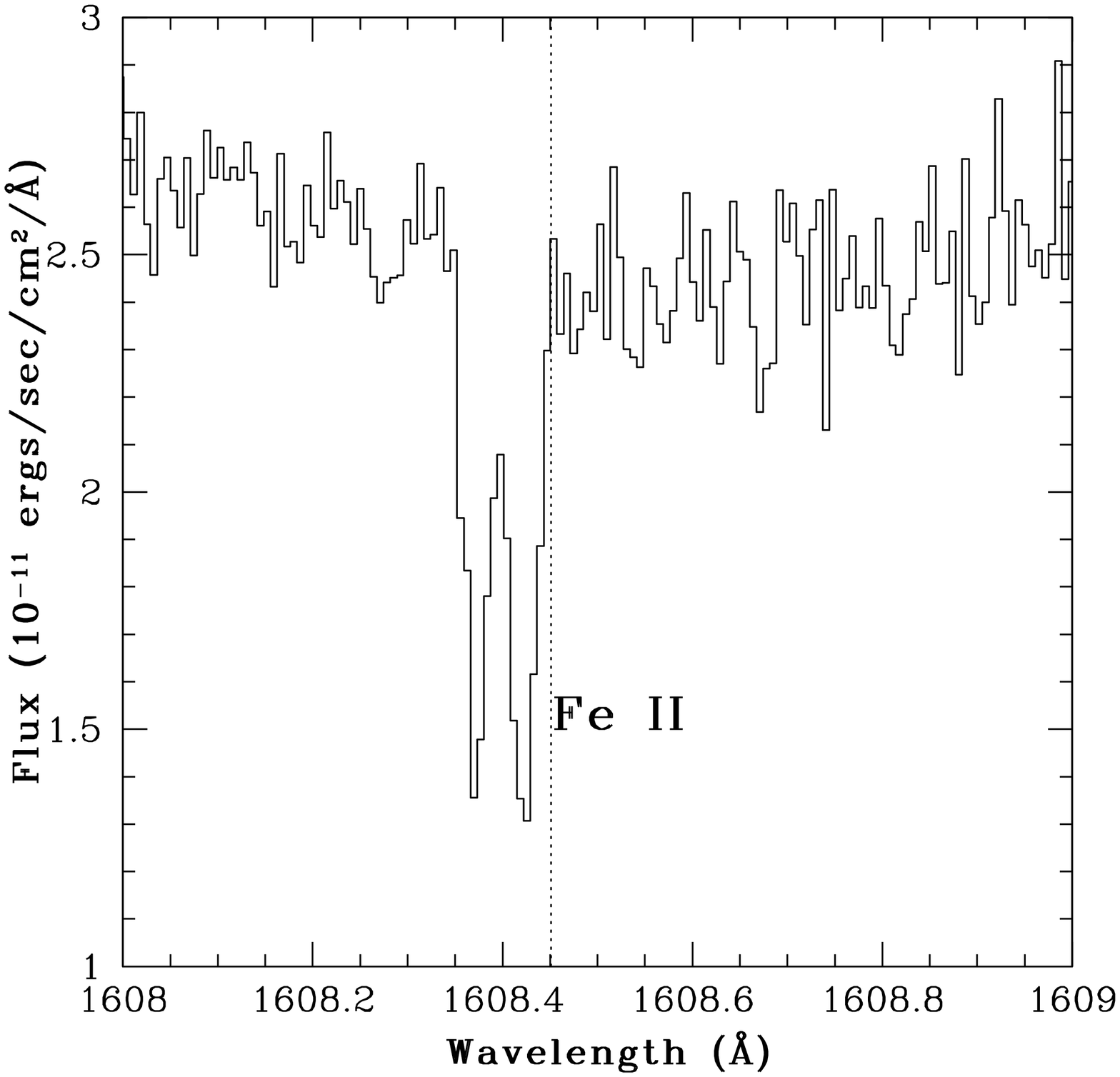}
\plotone{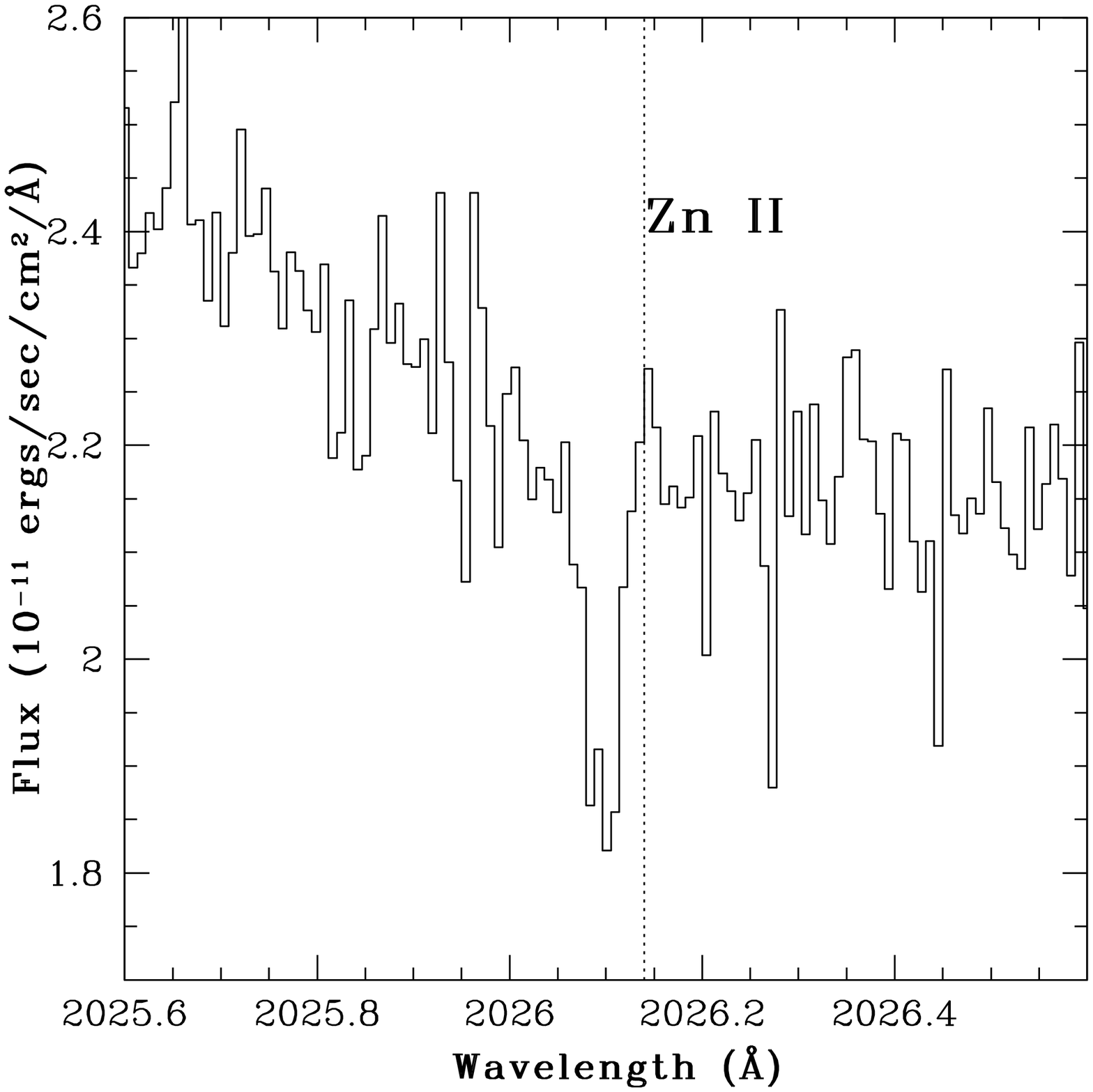}
\caption{The interstellar (a) \ion{Si}{2} $\lambda$ 1526.71 and (b) 
\ion{Fe}{2} $\lambda$ 1608.45, and (c) \ion{Zn}{2} $\lambda$ 2026.14 
lines obtained with STIS. The vertical lines indicate the wavelengths of the
\ion{Si}{2}, \ion{Fe}{2}, and \ion{Zn}{2} lines at the star's velocity..}
\end{figure}

\begin{figure}[ht]
\figurenum{2}
\epsscale{0.5}
\plotone{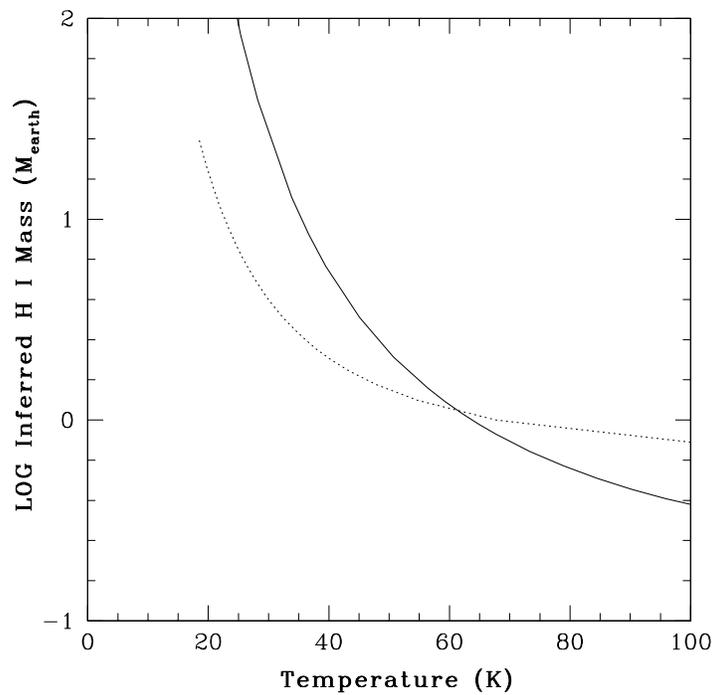}
\caption{The gas temperature, estimated from equation (3) plotted as a 
function of the gas:dust ratio or gas mass (dotted line). The circumstellar 
gas mass upper limit inferred from the \ion{Zn}{2} line at 2026.14 \AA \ 
plotted as a function of gas temperature (solid line). In these simple 
models, the gas has a constant temperature throughout the disk. Both 
relationships constrain the circumstellar gas mass as a function of 
temperature.}
\end{figure}

\begin{figure}
\figurenum{3}
\epsscale{1.0}
\plotone{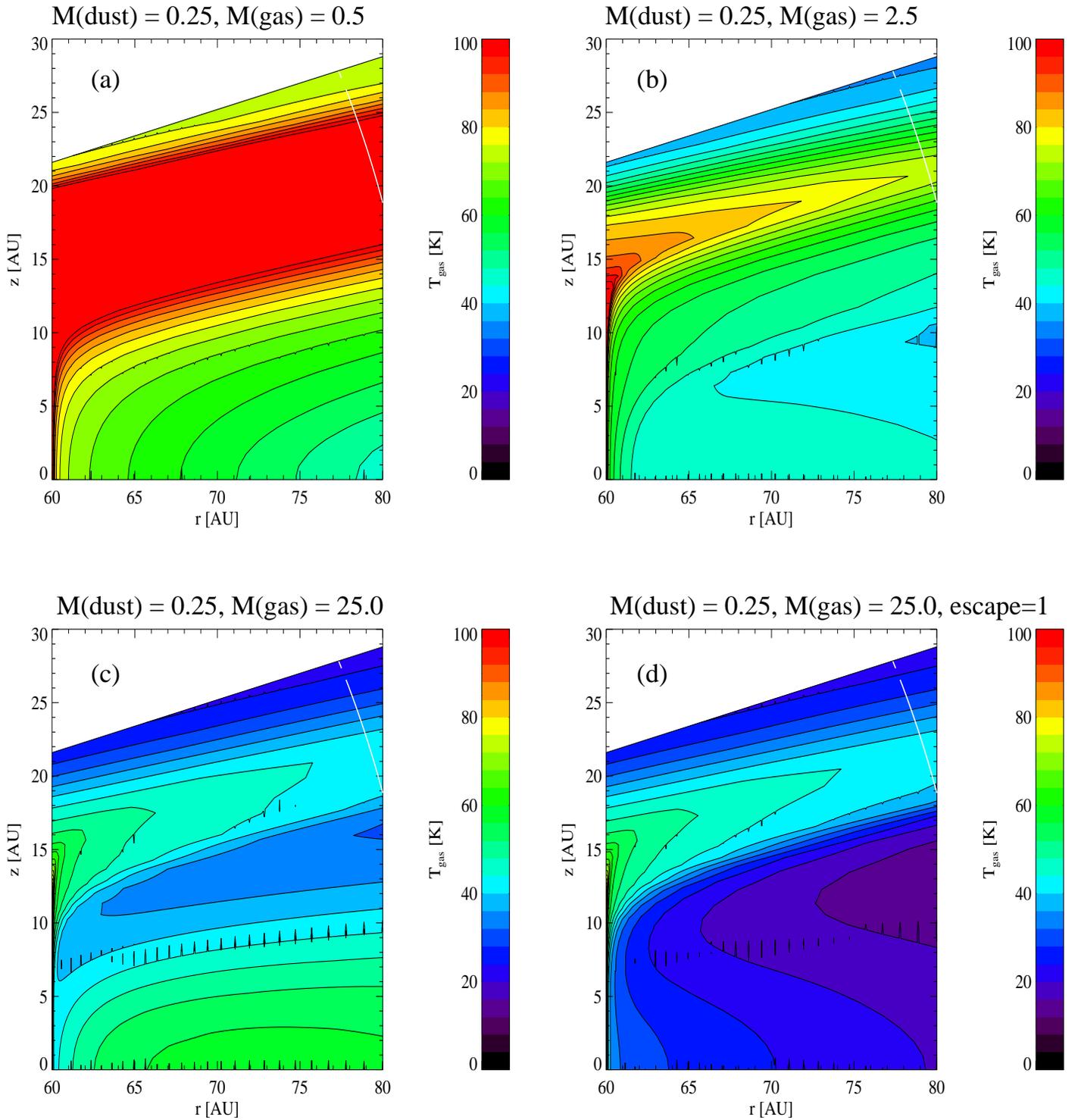}
\caption{(a)-(c) The gas temperature structure in a slice perpendicular to 
the disk for models with gas:dust ratios of 2, 10, and 100, calculated using 
the one dimensional escape probability formalism discussed in the text, and 
(d) the temperature structure for a model with a gas:dust ratio of 100 
calculated using $\beta$ = 1.0}
\end{figure}


\begin{thebibliography} {}
\bibitem[Abgrall et al. 1993]{abg93} Abgrall, H., Roueff, E., Launay, F.,
Roncin, J.-Y., \& Subtil, J.-L. 1993, \aaps, 101, 273
\bibitem[Augereau et al. 1999]{aug99} Augereau, J. C., Lagrange, A. M.,
Mouillet, D., Papaloizou, J. C. B., \& Grorod, P. A., \aap, 348, 557
\bibitem[Boss 2000]{bos00} Boss, A. 2000, \apj, 536, L101
\bibitem[Fajardo-Acosta, Telesco, \& Knacke 1998]{ftk98} Fajardo-Acosta,
S., Telesco, C. M., \& Knacke, R. F. 1998, \aj, 115, 2101
\bibitem[Gray 1992]{gray92} Gray, D. F. 1992, The Observation and Analysis
of Stellar Photospheres (Bristol: Cambridge University Press)
\bibitem[Greaves et al. 2000]{gre00} Greaves, J. S., Mannings, V., \& 
Holland, W. S. 2000, Icarus, 143, 155
\bibitem[Herczeg et a. 2002]{her02} Herczeg, G. J., Linsky, J. L., Valenti,
J. A., Johns-Krull, C. M., \& Wood, B. E. 2002, \apj, 572, 310
\bibitem[Hoffleit \& Warren 1991] {hw91} Hoffleit, D., \& Warren, W. H. 1991, 
The Bright Star Catalogue (5th ed.; New Haven: Yale Univ. Obs.)
\bibitem[Holweger, Hempel, \& Kamp 1999]{hhk99} Holweger, H., Hempel, M., \& 
Kamp, I. 1999, \aap, 350, 603
\bibitem[Jayawardhana et al. 1998]{jay98} Jayawardhana, R., Fisher, S.,
Hartmann, L., Telesco, C., Pina, R., \& Fazio, G. 1998, \apj, 503, L79
\bibitem[Jura et al. 1991]{jur91} Jura, M. 1991, \apj, 383, L79
\bibitem[Jura et al. 1995]{jur95} Jura, M., Ghez, A. M., White, R. J.,
McCarthy, D. W., Smith, R. C., \& Martin, P. G. 1995, \apj, 445, 451
\bibitem[Jura et al. 1998]{jur98} Jura, M., Malkan, M., White, R., Telesco, 
C., Pina, R., \& Fisher, R. S. 1998, \apj, 505, 897
\bibitem[Jura 2003]{jur03} Jura, M. 2003, to appear in in ASP Conf. Ser.,
Debris Disks and the Formation of Planets, eds. D. Backman, M. Meyer, \& L. 
Caroff (San Francisco: ASP)
\bibitem[Kamp \& van Zadelhoff 2001]{kvz01} Kamp, I. \& van Zadelhoff, G.-J.
2001, \aap, 373, 641 
\bibitem[Klahr \& Lin 2000]{kl00} Klahr, H., \& Lin, D. 2000, \apj, 554, 1095
\bibitem[Koerner et al. 1998]{koe98} Koerner, D. W., Ressler, M. E.,
Werner, M. W., \& Backman, D. E. 1998, \apj, 503, L83
\bibitem[Lecavelier des Etangs et al. 2001]{lec01} Lecavelier des Etangs, A.,
Vidal-Madjar, A., Roberge, A., Feldman, P. D., Deleuil, M., Andr\'{e}, M., 
Blair, W. P., Bouret, J.-C., et al. 2001, Nature, 412, 706
\bibitem[Li \& Lunine 2003]{ll03} Li, Aigen, \& Lunine, J. I. 2003, \apj,
590, 368
\bibitem[Liseau \& Artymowicz 1998]{la98} Liseau, R., \& Artymowicz, P. 
1998, \aap, 334, 935
\bibitem[Maloney, Hollenbach, \& Tielens 1996]{mht96} Maloney, P. R.,
Hollenbach, D. J., \& Tielens, A. G. G. M. 1996, \apj, 466, 561
\bibitem[Morton 1991]{dcm91} Morton, D. C. 1991, \apjs, 77, 119
\bibitem[Morton \& Noreau 1994]{mn94} Morton, D. C., \& Noreau, L. 1994,
\apjs, 95, 301
\bibitem[Richter et al. 2002]{ric02} Richter, M. J., Jaffe, D. T., Blake, 
G. A., \& Lacy, J. H. 2002, \apjl, 572, L161
\bibitem[Royer et al. 2002]{roy02} Royer, F., Grenier, S., Baylac, M.-O., 
Gomez A. E., Zorec, J. 2002, \aap, 393, 897
\bibitem[Ruden 1999]{rud99} Ruden, S. 1999, in The Origins of Stars
and Plantary Systems, eds. C. J. Lada and N. D. Kylafis (Dordrecht: Kluwer
Academic Publishers)
\bibitem[Sahnow et al. 2000]{sah00} Sahnow, D. J., Moos, H. W., Friedman, 
S. D., Blair, W. P., Conard, S. J., Kruk, J. W., Murphy, E. M., Oegerle, 
W. R. et al. 2000, SPIE, 4139, 131
\bibitem[Schneider et al. 1999]{sch99} Schneider, G., Smith, B. A., Becklin, 
E. E., Koerner, D. W., Meier, R., Hines, D. C., Lowrance, P. J., Terrile, 
R. J., Thompson, R. I., \& Rieke, M. 1999, \apj, 513, L127
\bibitem[Stauffer et al. 1995]{sta95} Stauffer, J. R., Hartmann, L. W., \&
Barrado y Navascu\'{e}s, D. 1995, \apj, 454, 910
\bibitem[Telesco et al. 2000]{tel00} Telesco, C. M., Fisher, R. S., Pina, 
R. K., Knacke, R. F., Dermott, S. F., Wyatt, M. C., Grogan, K., Holmes, 
E. K. et al. 2000, \apj, 530, 329
\bibitem[Thi et al. 2001]{thi01} Thi, W. F., Blake, G. A., van Dishoeck, 
E. F., van Zadelhoff, G. J., Horn, J. M. M., Becklin, E. E., Mannings, V.,
Sargent, A. I., et al. 2001, Nature, 409, 60
\bibitem[Tholen, Tejfel, \& Cox 2000]{ttc00} Tholen, D. J., Tejfel, V. G., 
\& Cox, A. N. 2000, in Allen's Astrophysical Quantities, ed. A. N. Cox
(New York: Springer-Verlag),  293
\bibitem[Tielens \& Hollenbach 1985]{th85} Tielens, A.G.G.M., Hollenbach, D.
1985, \apj, 291, 722
\bibitem[Weinberger, Becklin, \& Schneider 2000]{wbs00} Weinberger, A., 
Becklin, E., \& Schneider, G. 2000, in ASP Conf. Ser. 219, Disks, 
Planetesimals, and Planets, ed. F. Garzon, C. Eiroa, D. de Winter, \& T. J. 
Mahoney (San Francisco: ASP), 329
\bibitem[Wyatt et al. 1999]{wya99} Wyatt, M. C., Dermott, S. F., Telesco, 
C. M., Fisher, R. S., Grogan, K., Holmes, E. K., \& Pina, R. K. 1999, \apj
527, 918
\bibitem[York \& Jura 1982]{yj82} York, D. F., \& Jura, M. 1982, \apj,
254, 88
\bibitem[Zuckerman, Forveille, \& Kastner]{zfk95} Zuckerman, B., Forveille,
T. \& Kastner, J. 1995, Nature, 373, 494
\end{thebibliography}
\end{document}